\documentclass[10pt]{iopart}
\usepackage{iopams}
\usepackage{setstack}

\usepackage[english]{babel}
\usepackage{mathrsfs}
\usepackage{amsthm}
\usepackage{eucal} 
\usepackage[dvips]{graphicx} 
\usepackage[vcentermath]{youngtab}
\usepackage{verbatim}  
\usepackage[usenames]{color}

\newcommand{\defeq}{\doteqdot} \newcommand{\nat}{\mathbb{N}}
\newcommand{\reals}{\mathbb{R}} \newcommand{\cplx}{\mathbb{C}}
\newcommand{\proj}{\mathbb{P}}

\renewcommand{\textnormal}[1]{{\rm #1}}

\newcommand{\reno}{\mathscr{R}}

\newcommand{\perm}[1]{\mathfrak{S}_{#1}} 
 
 \newcommand{\deco}{\mathscr{D}}

\newcommand{\dominanti}[1]{\mathscr{M}_#1}
\newcommand{\holo}[1]{\mathscr{H}_#1}

\newcommand{\multiN}{\mathbf{\balpha}}
\newcommand{\decoedge}{\mathscr{E}}

\newcommand{\bauth}[1]{#1}
\newcommand{\btitle}[1]{#1} \newcommand{\bjou}[1]{\emph{#1}}
\newcommand{\bvol}[1]{\textbf{#1}} 
\newcommand{\bpage}[1]{#1} \newcommand{\byear}[1]{#1}
\newcommand{\bentry}[6]{\bauth{#1} \byear{#6} \btitle{#2} \bjou{#3}
  \bvol{#4}, \bpage{#5}}

\newcommand{\st}{\textnormal{\ s.t.\ }} \newcommand{\fa}{\forall\,}
\newcommand{\ex}{\exists\,} 
\newcommand{\deh}{\rmd} 
\newcommand{\ma}{\deh \deh ^c}

\newcommand{\fpart}{\mathscr{Z}} \newcommand{\hamil}{\mathscr{H}}

\newcommand{\enel}{\mathscr{F}} 

\newcommand{\fatou}{\mathcal{F}}
\newcommand{\julia}{\mathcal{J}}

\renewcommand{\star}{{\textcolor{Gray}{\rule[-0.4pt]{2.4pt}{2.4pt}}}}
\newcommand{\gstar}{{\textcolor{Gray}{\rule[-1.3pt]{7.3pt}{7.3pt}}}\vspace{-8pt}}

\newtheorem{obs}{Note}[section]
 \newtheorem{theorem}[obs]{Theorem}
 \newtheorem{proposition}[obs]{Proposition}
 
\newtheorem{definition}[obs]{Definition}
\newtheorem{example}[obs]{Example}
\newtheorem*{appdefinition}{Definition}
\newtheorem*{apptheorem}{Theorem}

\newenvironment{mydim}{\emph{Pf:}}{\hspace{\stretch{1}}$\blacksquare$\\[10pt]}

\newcommand{\dynspace}{\mathscr{M}}
\newcommand{\physspace}{\mathscr{P}}

\newcommand{\ri}[1]{{\alpha_{#1}}}
\newcommand{\rib}[1]{{\beta_{#1}}}
\newcommand{\ric}[1]{{\gamma_{#1}}}
\newcommand{\rid}[1]{{\delta_{#1}}}
\newcommand{\bw}{\mathcal{W}}
\newcommand{\bwp}{\tilde{\mathcal{W}}}
\newcommand{\Jp}{J_{\rm s}}
\newcommand{\Jnp}{J_{\rm d}}
\newcommand{\zp}{z_{\rm s}}
\newcommand{\znp}{z_{\rm d}}

\begin{document}
\title[Potts models on HLs and RG dynamics]{Potts models on hierarchical lattices and Renormalization Group dynamics}
\author{J De Simoi$^1$ and S Marmi$^2$}

\address{$^1$ University of Maryland, College Park, MD 20740, USA}
\address{$^2$ Scuola Normale Superiore, 56100 Pisa, Italy }
\ead{\mailto{jacopods@math.umd.edu}, \mailto{s.marmi@sns.it}}
\begin{abstract}
We prove that the generator of the renormalization group of Potts models on hierarchical lattices can be represented by a rational map acting on a finite-dimensional product of complex projective spaces. In this framework we can also consider models with an applied external magnetic field and multiple-spin interactions. We use recent results regarding iteration of rational maps in several complex variables to show that, for some class of hierarchical lattices, Lee-Yang and Fisher zeros belong to the unstable set of the renormalization map. 
\end{abstract}
\section{Introduction} 
Potts models on hierarchical lattices have been introduced in 1979 by Berker and Ostlund \cite{boh} as an interpretation of Migdal-Kadanoff models, defined in 1975 \cite{Mi1, Mi2, Kad} in order to approximate classical spin models on $\mathbb{Z}^d$. Later, in 1981, Griffiths and Kaufman \cite{kg1, kg3, kg2} provided a rigorous definition of hierarchical lattices and studied some examples in detail. One of such examples, the diamond hierarchical lattice, was later considered in a paper by Derrida, De Seze and Itzykson \cite{dsi}, who showed that the generator of the renormalization group (see e.g. \cite{fis, wil}) could be written as a rational map acting on the Riemann sphere \smash{$\hat\cplx$}; as a consequence, the Fisher set of the model coincides with the Julia (i.e. unstable) set of the renormalization group map. Later, similar results were established to study other specific lattices (e.g \cite{gasm, bcd}) or to introduce coupling with an external magnetic field in a similar dynamical framework (e.g. \cite{bly}).\\
In this paper we generalize the result of \cite{dsi} to all hierarchical lattices, i.e. we prove that the generator of the renormalization group of a Potts model on a hierarchical lattice can be represented by a rational map acting on a complex multiprojective space (sections \ref{hierarchical} and \ref{dynproj}). The general approach that we introduce, not only allows to describe all models on hierarchical lattice that have already been studied, but it also provides an extremely natural way to deal with an external magnetic field (\sref{physical}). The study of the dynamics obtained by iteration of a rational map in several complex variables is a quite recent research subject and, as such, it is still quite incomplete. Nevertheless, recent results by Dinh-Sibony \cite{ds} allow us to prove that, at least for some class of hierarchical lattices, Lee-Yang and Fisher sets are a subset of the Julia set of the renormalization map (\sref{main}).  
This paper features two technical appendices that give the basic mathematical background needed to understand the statements in the main part and provide references for the interested reader. A number of examples of Potts models on hierarchical lattices are presented in \cite{pap2}, where it is shown how to obtain both exact and numerical results by using the general methods developed in this paper. 
\section{Potts models on hierarchical lattices}\label{hierarchical}
In order to state our result in full generality, we need to provide formal definitions and notations for the objects we will use in the paper. In spite of the technical nature of such definitions, they are indeed quite natural and, most importantly, they will lead to a very simple proof of the result.
\subsection{Hypergraphs and hierarchical lattices}
\emph{Hierarchical lattices} (in short \emph{HL}s) are lattices that are left invariant by a given coarse-graining operation. The most famous example is provided by the diamond hierarchical lattice \cite{boh,kg3,dsi} which is obtained by iterating the substitution which replaces an edge with four edges linking the original vertices with two new (internal) vertices (see, e.g. \cite{kg3}, figure 1 or \cite{pap2}, figure 1). Our goal is to extend this procedure so as to be able to consider more general cases. To this purpose, we are going to define hierarchical lattices as limits of sequences of finite objects obtained iterating a \emph{decoration} procedure, which is going to be dual to the coarse-graining operation. The finite objects we consider are a generalization of graph called \emph{hypergraph} (see e.g. \cite{hyp}); hypergraphs have been briefly considered in \cite{kg3} (see section V) under the name of ``generalized graphs'' and they were used for defining hierarchical lattices with multiple spin interactions. In fact, hypergraphs differ from graphs in the sense that edges (sometimes also called \emph{hyperedges} or \emph{links}) are allowed to connect an arbitrary number of vertices. Hereby follows the standard definition.
\begin{definition}
A hypergraph $\Gamma$ is defined by a set $V$ of vertices and a set $E$ of edges that are finite ordered non-empty subsets of $V$; the same vertex cannot appear more than once in an edge. Given an edge $e$, we define \emph{rank} of $e$ its cardinality $|e|$ as a subset of $V$. If all edges have the same rank $r$, the hypergraph is said to be $r-$uniform and $r$ is said to be the \emph{order} of the hypergraph.\\
Given a hypergraph $\Gamma=\{V,E\}$, a \emph{partial hypergraph} $\Gamma'=\{V',E'\}\subset\Gamma$ is defined as a hypergraph such that $V'=V$ and $E'\subset E$.
\end{definition}
From the physical point of view, edges will connect spins that are coupled to each other; notice that the definition only takes into account edges of finite rank as we do not consider interactions of infinite range. We \emph{do not} assume that either $V$ or $E$ are finite.\\
It will be necessary to associate different properties (e.g. interactions) to edges of the same rank; such properties will be indexed by a (at most) countable index set $\mathcal{I}$ (the set of \emph{types}) that will be common among all hypergraphs; the notion of \emph{structured hypergraph} takes into account this additional piece of information.
In order to define it, we first need to introduce the notion of \emph{partition} of a hypergraph $\Gamma=\{V,E\}$ \emph{into uniform partial hypergraphs} $\Gamma_{(r,i)}$, where the \emph{rank} $r\in\nat$ and the \emph{type} $i\in\mathcal{I}$ have been fixed. This partition is obtained as follows: we define $E_{(r,i)}$ to be the set of all edges of $\Gamma$ with rank equal to $r$ and type equal to $i$. One of course has: 
\[E_{(r,i)}\cap E_{(s,j)}=\emptyset\textnormal{\ if\ }r\not =s\textnormal{\ or\ } i\not =j.
\]
The edge set $E$ of the original hypergraph will be the disjoint union
\[
E=\bigsqcup E_{(r,i)}
\]
and, denoting with $\Gamma_{(r,i)}$ the uniform partial hypergraph $\Gamma_{(r,i)}=\{V,E_{(r,i)}\}$, we have
\[
\Gamma=\bigcup_{(r,i)\in\nat\times\mathcal{I}}\Gamma_{(r,i)}.
\]
Note that since each $\Gamma_{(r,i)}$ is $r$-uniform, each element of $E_{(r,i)}$ is an ordered $r-$tuple of vertices. The space of all pairs (rank,type) is called $\mathscr{A}\defeq\nat\times\mathcal{I}$ and we denote its elements by Greek letters e.g. $\alpha=(r,i)$.\\
We can now define a \emph{structured hypergraph} $\bGamma$ as a hypergraph $\Gamma$ along with a partition into 
 uniform partial hypergraphs.  The sets $E_\ri{}$ will be called \emph{partial edge sets} of $\bGamma$. We define the \emph{multiorder} of $\bGamma$ to be the set $\multiN\defeq\{\ri{}\in\mathscr{A} \st E_\ri{}\not =\emptyset\}$. When $\multiN=\{\ri{1},\cdots,\ri{p}\}$ is \emph{finite}, then $\bGamma$ is said to be a \emph{finitely structured hypergraph}; if $\multiN=\{\ri{}\}$ has only one element (i.e. $p=1$), then $\bGamma$ is said to be $\ri{}-$uniform and $\ri{}$ will be called \emph{order} of the structured hypergraph. 
Given a structured hypergraph $\bGamma=\bigcup_\ri{}\Gamma_\ri{}$, it is convenient to consider each uniform partial hypergraph $\Gamma_\ri{}$ also as a $\ri{}-$uniform structured hypergraph $\bGamma_\ri{}$, so that we can write $\bGamma=\bigcup_\ri{}\bGamma_\ri{}$.    
\begin{figure}[h!] 
\begin{center}
\includegraphics[width=7cm]{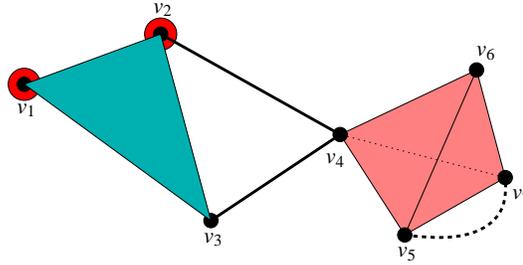}
\end{center}
\caption{An example of non-uniform structured hypergraph ($p=5$)}
      \begin{equation*}
        \begin{array}{lcll}
V&=&\left\{v_1,v_2,v_3,v_4,v_5,v_6,v_7\right\}&\\[3pt]
E_{(1,1)}&=&\left\{(v_1),(v_2)\right\}&\mathrm{rank\ 1\ type\ 1}\\
E_{(2,1)}&=&\left\{(v_2,v_4),(v_3,v_4)\right\}&\mathrm{rank\ 2\ type\ 1}\\
E_{(2,2)}&=&\left\{(v_5,v_7)\right\}&\mathrm{rank\ 2\ type\ 2}\\
E_{(3,1)}&=&\left\{(v_1,v_2,v_3)\right\}&\mathrm{rank\ 3\ type\ 1}\\
E_{(4,1)}&=&\left\{(v_4,v_5,v_6,v_7)\right\}& \mathrm{rank\ 4\ type\ 1}\\
        \end{array}
      \end{equation*} 
\begin{equation*}
\bGamma=\{V,E=E_{(1,1)}\sqcup E_{(2,1)}\sqcup E_{(2,2)}\sqcup E_{(3,1)} \sqcup E_{(4,1)} \}
\end{equation*}
\label{hypergraph}
\end{figure}  
For convenience of notation, in the remaining of this section we will consider only \emph{finitely structured hypergraphs}; all the statements can easily be generalized to the infinite case. 
Let $\bGamma$ be a structured hypergraph $\bGamma=\{V,E=E_\ri{1}\sqcup\cdots\sqcup E_\ri{p}\}$ and let $f$ be a map $f$ from the set $V$ to another set $W$ such that the restriction of $f$ to every edge $e\in E$ is injective; we call such an $f$ a \emph{locally injective map}. Given a locally injective map $f$, for all $\ri{}=(r,i)$ we can induce a map $f_*$ from each $E_\ri{}$ to the set of ordered $r$-tuples of $W$ as follows:
\[
f_* \big(e=(v_1,\cdots,v_{r})\big)=\big(f(v_1),\cdots,f(v_{r})\big).
\] 
By local injectivity, $f_*E$ can be regarded as an edge set on $W$ and we can define (with a slight abuse of notation) $f_*\bGamma=\{W,f_*E=f_*E_\ri{1}\sqcup\cdots\sqcup f_*E_\ri{p}\}$ as the \emph{structured hypergraph induced by} $f$.  
The decoration procedure we want to define (that will be dual to the coarse-graining operation) will consist in gluing a fixed structured hypergraph to each edge of a given rank and type of another structured hypergraph. In order to do so, we need to mark the vertices which will be used in the gluing process: structured hypergraphs with marked vertices will be called \emph{decorated edges} (see e.g. figure \ref{spiderweb}).
\begin{definition}
Let $\ri{}=(r,i)$. A \emph{decorated $\ri{}-$edge} $\decoedge$ (of rank $r$ and type $i$) is a structured hypergraph $\bGamma$ with $r$ marked vertices.
\end{definition}
Marking vertices amounts to choose an additional \emph{ordered} $r-$tuple of vertices; borrowing the terminology from \cite{kg3}, section V, marked vertices will be called \emph{external (or surface) vertices} and vertices that are not external will be called \emph{internal (or core) vertices}.
\[
\decoedge=\{V=(v_1,\cdots,v_r)\sqcup V_0,E= E_\rib{1} \sqcup \cdots \sqcup E_\rib{p}\}.
\]
A decorated $\ri{}-$edge is said to be uniform if the underlying hypergraph $\bGamma$ is $\ri{}-$uniform.\\

\begin{figure}[h!]
\begin{center}
\includegraphics[width=3cm]{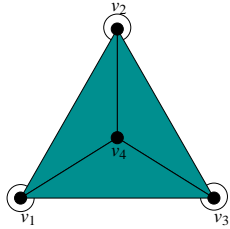}
\caption{An example of uniform decorated edge of rank 3. External vertices are circled.}
      \begin{equation*}
        \begin{array}{lcl}
V&=&\left\{(v_1,v_2,v_3)\sqcup \{v_4\}\right\}\\[3pt]
E_{(3,1)}&=&\left\{(v_1,v_4,v_3),(v_2,v_4,v_1),(v_3,v_4,v_2)\right\}
        \end{array}
      \end{equation*}  
\begin{equation*}
\bGamma=\{V,E= E_{(3,1)} \}
\end{equation*}
\label{spiderweb} 
\end{center} 
\end{figure}
Decorated edges can be physically regarded as the inner structure of an edge of a given rank and type. Notice moreover that the value of $i$ is not taken into consideration in the definition of a general decorated edge; it will, however, play a role in what follows. 
It is easy (see \ref{sum}) to introduce a natural notion of sum on decorated $\ri{}-$edges; the attempt to define a natural multiplication operation leads to a fundamental operation on a structured hypergraph $\bGamma$ which will be called \emph{decoration}. It amounts to substituting edges of rank $r$ and type $i$ in a hypergraph with given decorated edges of the same rank and type.\\
Let $\ri{}=(r,i)$ be fixed, $\bGamma=\{V,E=E_\ri{}\}$ be an $\ri{}-$uniform structured hypergraph and $\decoedge=\{W=(w_1,\cdots,w_r)\sqcup W_0,F=F_\rib{1}\sqcup\cdots\sqcup F_\rib{p}\}$ be a decorated $\ri{}-$edge. The \emph{product} of $\bGamma$ with $\decoedge$ is the structured hypergraph given by the following procedure: each edge $e\in E$ is removed from $\bGamma$ and replaced by a copy of $\decoedge$, with surface vertices of $\decoedge$ identified to the vertices of $e$, respecting their ordering. The partition in uniform partial hypergraphs for the resulting  hypergraph will be the one induced by the partition of $\decoedge$; the resulting structured hypergraph will be denoted by $\bGamma\times\decoedge$: more formally, let $\tilde V \defeq V\sqcup E\times W_0$. If we define the collapsing map $\pi$ as follows:
\[ \pi:E\times W \to\tilde V,\]\[\pi\left(e=(v_1,\cdots,v_r),w\right)=
\cases{v_l&if  $w=w_l$ for some $l$\\
     (e,w)&otherwise}
,
\]
then the edge sets are given by:
\[
\tilde E_\rib{}\defeq\pi_* ( E\times F_\rib{}),\ \rib{}\in\bbeta=\{\rib{1},\cdots,\rib{p}\},
\]
and the resulting structured hypergraph will be
\[\bGamma\times\decoedge\defeq\{\tilde V, \tilde E=\tilde E_\rib{1} \sqcup\cdots\sqcup\tilde E_\rib{p}\}.
\]
Given a structured hypergraph $\bGamma$, one can multiply simultaneously and independently each $\ri{}-$uniform partial hypergraph of the partition $\bGamma=\bigcup\bGamma_\ri{}$ with a decorated $\ri{}-$edge $\decoedge_\ri{}$.\\
We define the \emph{identity} decorated $\ri{}-$edge to be the uniform decorated $\ri{}-$edge with $r$ surface vertices, no core vertices and only one $\ri{}-$edge (of rank $r$ and type $i$) connecting the surface vertices with the correct ordering.
\[
1_\ri{}\defeq\{V=(v_1,\cdots,v_r)\sqcup\emptyset,E=E_\ri{}=\{(v_1,\cdots,v_r)\}\}.
\]
\begin{definition}
We define a \emph{decoration} $\deco$ as a choice of decorated edges $\{\decoedge_\ri{}\}_{\ri{}\in\mathscr{A}}$, such that only finitely many $\decoedge_\ri{}$ are different from $1_\ri{}$. Then $\deco$ acts on a structured hypergraph $\bGamma$ as follows:
\[
\deco_{\{\decoedge_\ri{}\}}\bGamma=\bigcup_\ri{}\big(\bGamma_\ri{}\times\decoedge_\ri{}\big).
\]          
\end{definition}
Notice that if we choose $\decoedge_\ri{}=1_\ri{}$ for all $\ri{}\in\mathscr{A}$ we have the identity operation $\deco_{\{1_\ri{}\}}\bGamma=\bGamma$. For notational convenience we will explicitly write as subscripts of $\deco$ only the non-trivial decorated edges involved in the decoration procedure. Moreover, it is clear that any decoration of a finitely structured hypergraph with decorated edges that are themselves finitely structured will yield a finitely structured hypergraph.
Note that we can define a decoration operation in the class of decorated $\ri{}-$edges by applying the decoration to the underlying structured hypergraph and keeping the same external vertices. This last remark allows us to define the composition of decorations in the following natural way: let $\deco_1=\deco_{\{\decoedge_1,\cdots,\decoedge_k\}}$ and $\deco_2$ be two decorations; then we define the composite decoration as:  
\[\deco_2\deco_1\defeq\deco_{\{\deco_2\decoedge_1,\cdots,\deco_2\decoedge_k\}}.\]\

\begin{figure}[h!]
\begin{center}
\includegraphics[width=8cm]{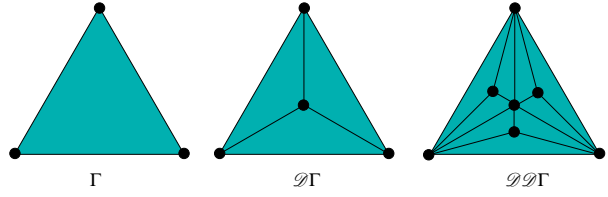} 
\caption{Decorating the structured hypergraph $\bGamma$ with the decorated edge of \fref{spiderweb} }
\label{iterations}
\end{center} 
\end{figure}

Using the decoration procedure, we introduce a partial ordering in the class of structured hypergraphs. We say that $\bGamma_1\leq\bGamma_2$ if there exists a decoration procedure $\deco$ such that $\deco\bGamma_1=\bGamma_2$. \\
We now fix a a decoration operation $\deco$ and a finite initial hypergraph $\bGamma_0$; decorating $\bGamma_0$ will yield $\bGamma_1=\deco\bGamma_0$. If we iterate the action of $\deco$ (see e.g. \fref{iterations}) there can be two cases: either at some point the decoration operation acts trivially on the obtained hypergraph because we run out of edge to decorate, or not. The former case corresponds to \emph{finitely renormalizable lattices}; in the latter case the infinite lattice $\bGamma_\infty$ obtained as the inductive limit of the decoration procedure is called a \emph{hierarchical lattice}. 
In this setting it is now clear that the decoration operation is dual to the coarse-graining process that amounts to gluing the original edges back in the place of the corresponding decorated edges. Moreover, notice that this is only one particular way to construct an infinite lattice using decorations; for instance it would be interesting to study thermodynamical properties of an infinite lattice obtained by fixing two (or more) decorations and then choosing one or the other at random to define the sequence $\bGamma_n$ (i.e. a random walk on decorations).    
\\[3pt]
In all subsequent sections we will only deal with finitely structured hypergraphs, therefore, without risk of confusion, we will drop the words ``finitely structured'' and use just the word ``hypergraph''. 
\subsection{Interactions on hierarchical lattices. Potts models.}
We will consider Potts models on hierarchical lattices; Hamiltonians will be obtained by summing over all edges a local interaction that depends only on the states of the spins belonging to the edge, i.e. a nearest-neighbours interaction. It is worthwhile to notice that, since edges of hypergraphs may connect an arbitrary number of vertices, such interactions are not restricted to pair interactions; this flexibility turns out to be useful as, for instance, it allows at the same time to deal with external magnetic fields (by considering edges of rank 1) or to study the more complicated interactions that arise renormalizing a pair interaction.\\
Let $q\geq 2$ be the number of Potts states of the model; for a given hypergraph ${\bGamma}=(V,E)$, a \emph{configuration} $\sigma$ is a map from $V$ to $S\defeq\{1,\cdots,q\}$. In order to associate an energy to each configuration we,  first need to fix the nearest-neighbours interactions: this amounts, for each edge set $E_\ri{},\,\ri{}=(r,i)$, to fix the energy contribution of the configuration of the $r$ spins connected by such edges, i.e. to fix $q^{r}$ complex numbers. Such numbers will be denoted by $J^\ri{}_I=J^\ri{}_{s_1\cdots s_{r}}$, where $s_k\in S$ and $I$ is a multi-index ranging over $S^{r}$. The total energy associated to a configuration $\sigma$ is therefore easily expressed in terms of such $J^\ri{}_I$:
\[
\hamil^{{\bGamma}}(\sigma)=\sum_{\ri{}\in\balpha}\ \sum_{(v_1,\cdots,v_{r})\in E_\ri{}}J^\ri{}_{\sigma(v_1)\cdots\sigma(v_{r})}.
\]
The associated partition function is:
\[
\fpart^{\bGamma}=\sum_{\sigma\in S^V}\exp\left(-\beta\hamil^{\bGamma}(\sigma)\right),
\]
where $\beta=1/kT$; define now the \emph{Boltzmann weights} as:
\[
\fa\ri{}=(r,i),\ z^\ri{}_I\defeq \exp\left(-\beta J^\ri{}_I\right);\quad z^\ri{}\in\bw^\ri{}\defeq\cplx^{q^r}.
\]
In such coordinates, each term $\exp\left(-\beta\hamil^{\bGamma}(\sigma)\right)$ is a monomial of degree given by the number of edges in the hypergraph. If we fix a partial edge set $E_{\bar\ri{}}$, the degree of the polynomial in the variables $z^{\bar\ri{}}_I$ is given by the number of edges in $E_{\bar\ri{}}$. Thus, $\fpart^{\bGamma}$ is a \emph{homogeneous polynomial} that is \emph{separately} homogeneous in $z^\ri{}_I$ for all fixed $\ri{}$. As decorated edges of rank $r$ are hypergraphs with $r$  marked vertices, it is natural to consider the \emph{conditional} partition functions of a decorated edge, for which we specify the $r$ states $(s_1,\cdots,s_r)$ of the external vertices $(v_1,\cdots,v_r)$ and restrict the sum to configurations satisfying the condition: 
\[
\fpart_{s_1\cdots s_r}^\decoedge \defeq\sum_{\substack{\sigma\in S^V \cr \sigma(v_k)=s_k\ k=1,\cdots,r}} \exp\left(-\beta\hamil^\decoedge(\sigma)\right).
\] 
Once more, these are homogeneous and separately homogeneous polynomials in $z^\ri{}_I$ of fixed degree, independent of the choice of the external states. It is easy to check that the identity edge $1_\ri{}$ gives the trivial conditional partition function $\fpart^{1_\ri{}}_I=z^\ri{}_I$.
\subsection{The renormalization map}
 Conditional partition functions provide a natural way to connect the partition function of a hypergraph and the partition function of its image under decorations.
\begin{definition}
Consider a decorated $\ri{}-$edge $\decoedge=\{W=(w_1,\cdots,w_r)\sqcup W_0,F=F_\rib{1}\sqcup\cdots\sqcup F_\rib{p}\}$. We define the renormalization map
\[
\reno^\decoedge: \bw^\rib{1}\times\cdots\times\bw^\rib{p}\to\bw^\ri{},\]
as given in coordinates by the conditional partition functions: 
\[
\big(\reno^\decoedge(z^\rib{1},\cdots,z^\rib{p})\big)_I= \fpart^\decoedge_I(z^\rib{1},\cdots,z^\rib{p})
\]
\end{definition}
Consider a $\ri{}-$uniform hypergraph ${\bGamma}$; the partition function of ${\bGamma}$ is the polynomial $\fpart^{\bGamma}:\bw^\ri{}\to\cplx$. If we multiply ${\bGamma}$ with $\decoedge$ we obtain a hypergraph ${\bGamma}\times\decoedge$ whose partition function is the polynomial $\fpart^{{\bGamma}\times\decoedge}: \bw^\rib{1}\times\cdots\times\bw^\rib{p}\to\cplx$. The fundamental property of the partition function of a product is that it is obtained by composing the original partition function with the renormalization map, i.e. we claim that: 
\[
\fpart^{{\bGamma}\times\decoedge}(z^\rib{1},\cdots,z^\rib{p})=\fpart^{\bGamma}\circ\reno^\decoedge(z^\rib{1},\cdots,z^\rib{p})
\]
In fact, one can rewrite the sum over configurations involved in the partition function of ${\bGamma}\times\decoedge$ by \emph{first} summing over the configurations of vertices that belong to ${\bGamma}$ as well, \emph{then} over configurations of all vertices that have been generated by decorating each edge of ${\bGamma}$. In this way it is straightforward to see that $\fpart^{{\bGamma}\times\decoedge}$ is obtained by substituting each occurrence of $z_I$ in $\fpart^{\bGamma}$ with $\fpart^\decoedge_I$.\\[3pt]
A decoration $\deco$ amounts to a choice, for all $\ri{}\in\mathscr{A}$, of a decorated $\ri{}-$edge $\decoedge_\ri{}$ such that only finitely many $\decoedge_\ri{}$ are different from the identity.
 To each $\decoedge_\ri{}$ we can associate its renormalization map: 
\[
\reno^{\decoedge_\ri{}}: \bw^\rib{\ri{},1}\times\cdots\times\bw^\rib{\ri{},p_\ri{}}\to\bw^\ri{},\]
and finally we can define the renormalization map $\reno^\deco$ as the juxtaposition of the maps $\reno^{\decoedge_\ri{}}$, i.e.:
\[
\reno^\deco:\prod_{\ri{}\in\mathscr{A}}\bw^\ri{}\to\prod_{\ri{}\in\mathscr{A}}\bw^\ri{}  \qquad \pi_\rib{}\reno^\deco = \reno^{\decoedge_\rib{}}.
\]
where $\pi_\rib{}:\prod_{\ri{}}\bw^\ri{}\to\bw^\rib{}$ is the natural projection. \\
Now consider the case of general hypergraphs; let \protect{${\bGamma}=\{V,E=E_\ri{1}\sqcup\cdots\sqcup E_\ri{p}\}$} be a structured hypergraph; its partition function is a polynomial $\fpart^{\bGamma}:\bw^\ri{1}\times\cdots\times\bw^{\ri{p}}\to\cplx$. Let  \protect{${\bGamma}'=\deco_{\{\decoedge_\ri{}\}}{\bGamma}$}; the partition function of ${\bGamma}'$ is a polynomial \smash{$\fpart^{{\bGamma}'}:\bw^{\rib{1}}\times\cdots\times\bw^{\rib{q}}\to\cplx$}.
Again the claim is:
\begin{equation}
\fpart^{{\bGamma}'}(z^\rib{1},\cdots,z^\rib{q})=\fpart^{\bGamma}\circ\reno^\deco(z^\rib{1},\cdots,z^\rib{q})
\label{unicum}
\end{equation}
and it follows by applying the previous argument to each element of the partition into partial uniform hypergraphs.\\[9pt]
The relation between the decoration operation $\deco$ and the renormalization map $\reno^\deco$ is contravariant, i.e.:
\[\reno^{\deco_2\deco_1}=\reno^{\deco_1}\circ\reno^{\deco_2}.\]
In fact, the renormalization operation is covariant to the coarse-graining operation dual to the decoration procedure. Moreover, notice that the domain of the renormalization map $\reno^\deco$ is the infinite dimensional space of all interactions; however, since $\deco$ acts as the identity on all but finitely many edge sets, $\reno^\deco$ acts non-trivially on a finite dimensional space only. If we have a hierarchical lattice ${\bGamma}_\infty$ generated by the iteration of decoration procedure $\deco$, then $\reno^\deco$ can be iterated on the space of Boltzmann weights of ${\bGamma}_\infty$ and this space will be a finite dimensional complex vector space. As we will see later, the dynamics of $\reno^\deco$ will reflect thermodynamical properties of the Potts model on ${\bGamma}_\infty$. 
\section{The dynamical space: symmetries and interactions}\label{dynproj}
When defining the interactions $J^\ri{}$, we can choose the zero of energy for each edge set independently and arbitrarily. This freedom is reflected by the fact that the physics of the system will not change if we apply the map $J^\ri{}_{I}\mapsto J^\ri{}_{I}+\Delta^\ri{}$ or, equivalently, \smash{$z^\ri{}_{I}\mapsto z^\ri{}_{I}\cdot \exp\left(-\beta\Delta^\ri{}\right)$}, for an arbitrary choice of $\Delta^\ri{}$. This elementary observation allows us to establish an equivalence relation on each space of Boltzmann weights $\bw^\ri{}$ i.e.:
\[
z^\ri{},w^\ri{}\in\bw^\ri{},\quad z^\ri{}\sim w^\ri{}\ \mathrm{if\ }\ex\lambda\in\cplx\setminus\{0\}\st z^\ri{}_I=\lambda w^\ri{}_I\ \fa I;
\] 
equivalent Boltzmann weights will give identical physical systems.
If we take the quotient of $\bw^\ri{}=\cplx^{q^r}$ with respect to this equivalence relation, we obtain a \emph{projective space} \smash{$\bwp^\ri{}\defeq\proj^{q^r-1}$}. Thus, the quotient of the space of all Boltzmann weights with respect to all such equivalence relations is a product of projective spaces, i.e. a \emph{multiprojective space}, that will be called \emph{dynamical space} and will be denoted by $\dynspace$. Given ${\bGamma}=\{V,E=E_\ri{1}\sqcup\cdots\sqcup E_\ri{p}\}$, the dynamical space associated to ${\bGamma}$ will be the finite dimensional multiprojective space:
\[
\dynspace^{\bGamma}\defeq\bwp^\ri{1}\times\cdots\bwp^\ri{p}.
\]
Notice that, if we have a $\ri{}-$uniform hypergraph $\bGamma$, $\ri{}=(r,i)$, the dynamical space $\dynspace^{\bGamma}$ is a standard complex projective space of dimension $q^r-1$. Hereafter the Boltzmann weights $\{z\}$ will be considered to belong to the dynamical space and they will be denoted by $[z]$. Natural coordinates on the resulting projective space are \emph{homogeneous coordinates} of which we recall the definition in \ref{appA}.\\[3pt]
Notice that the renormalization map is a well-defined rational map on the dynamical space, since each coordinate given by a separately homogeneous polynomial. Moreover, the dynamical space of a hierarchical lattice is finite dimensional and invariant under the renormalization map. This means that at most a finite number of new interactions will be generated by the renormalization procedure; in this sense, Potts models on hierarchical lattices are completely renormalizable.
The approach we just presented is particularly convenient for studying the dynamics of the renormalization map, as the dynamical space has now been compactified in a natural way. All homogeneous thermodynamical quantities (e.g. susceptibility) can still be defined using variables in the dynamical space, but in order to define inhomogeneous quantities (such as free energy) we need to fix a zero of energy i.e. to consider variables belonging to the linear (not the projective) spaces. \\[3pt]
We will now look for invariant (projective) subspaces of the dynamical space; studying the dynamics of the renormalization map in such smaller subspaces is both interesting, as they correspond to special physically symmetric configurations, and convenient, as a map on a lower dimensional space is generally easier to study.\\
We are going to consider two different symmetries of the dynamical space: the first one is generated by $\perm{q}$, the group of permutations of $S$; the second symmetry is generated by the groups $\{\perm{}^{\ri{}}\}$, where each $\perm{}^\ri{}$ is the group of permutations of vertices of edges belonging to $E_\ri{}$.\\
The group $\perm{q}$ acts on the dynamical space in the following natural way:
\begin{definition}
Let $U\in\perm{q}$; for all $\ri{}$ we denote by $U^*$ the map $U^*:\bwp^{\ri{}}\to \bwp^{\ri{}}$ defined as follows:
\[ U^*\left(\left[z^\ri{}_{s_1\cdots s_{r}}\right]\right)= \left[z^\ri{}_{Us_1\cdots Us_{r}}\right].
\]
With a slight abuse of notation we denote by $U^*$ also the map that acts on an arbitrary product $\bwp^{\ri{1}}\times\cdots\times\bwp^{\ri{p}} $ by applying $U^*$ to each factor $\bwp^{\ri{k}}$.
\end{definition} 
The following proposition can be regarded as a general statement about the fact that if we perform the renormalization of a system with no external magnetic field, then the renormalized system will have no external magnetic field. More precisely: 
\begin{proposition}
For all $\decoedge$, the action of $\perm{q}$ commutes with $\reno^\decoedge$.
\label{white}
\end{proposition}
\begin{mydim}
By definition, each component of $\reno^\decoedge$ is a conditional partition function; let us consider the partition function associated to the choice of a multi-index $I$: 
\begin{equation*}
\fpart^\decoedge_I\left(\left[z^\ri{}_J\right]\right)=
\sum_{\begin{subarray}{c}\sigma\in S^V s.t. \cr \sigma(\mathrm{ext})=I\end{subarray}}\exp(-\beta\hamil^\decoedge(\sigma)).
\end{equation*}
Given an element $U\in\perm{q}$, we can write its action after the renormalization map:
\begin{eqnarray*}
\big(U^*\reno^\decoedge\big)_I\defeq\fpart^\decoedge_{UI}&=\sum_{\substack{\sigma\in S^V s.t \cr \sigma(\mathrm{ext})=UI}}\exp(-\beta\hamil^\decoedge(\sigma))=\\ &=\sum_{\substack{\sigma\in S^V s.t. \cr U^{-1}\sigma(\mathrm{ext})=I}}\exp(-\beta\hamil^\decoedge(\sigma)).
\end{eqnarray*}
Since the sum is over all the configuration space we can as well sum over $\sigma'\defeq U^{-1}\sigma$, so that:
\[
\big(U^*\reno^\decoedge\big)_I=\sum_{\substack{\sigma'\in S^V \st \cr \sigma'(\mathrm{ext})=I} }
\exp(-\beta\hamil^\decoedge(U\sigma'))=\fpart^\decoedge_I\left(\left[z^\ri{}_{UJ}\right]\right)\defeq\big(\reno^\decoedge U^*\big)_I.
\]
\end{mydim}
In all cases of interest, we will consider the action of a subgroup $G$ of $\perm{q}$ that is either going to be the whole group $\perm{q}$ (no external magnetic field: all states are considered equal) or $\perm{q-1}$ (external magnetic field: one state is special, all others are equal). Consider the subset of $\dynspace$ of points fixed by the action of $G$; then, proposition \ref{white} states that this subset is \emph{invariant} under $\reno^\deco$. This subset will turn out to be a lower dimensional multiprojective space naturally embedded in $\dynspace$. We will provide this embedding shortly, but first we need to describe the action of the other symmetry group.\\ 
Each group $\perm{}^{\ri{}}$ acts on the dynamical space in a natural way as well:
\begin{definition}
Let $V\in \perm{}^\ri{}$. We denote by $V^*$ the map $V^*:\left[z^\ri{}_{s_1\cdots s_{r}}\right]\mapsto\left[z^\ri{}_{s_{V1}\cdots s_{Vr}}\right]$ on the dynamical space. 
\end{definition}
Given a decorated $\ri{}-$edge $\decoedge$, $\reno^\decoedge$ does not necessarily commute with the action of $\perm{}^\ri{}$, since $\decoedge$ may have some internal structure that could break the symmetry. This amounts to say that if we renormalize a completely $\perm{}^\ri{}$-symmetric interaction we can possibly obtain a renormalized interaction that is \emph{not} $\perm{}^\ri{}$ symmetric. In fact, given a subgroup $H$ of $\perm{}^\ri{}$ we say that a decorated $\ri{}-$edge $\decoedge$ is $H-$symmetric if $\reno^\decoedge$ commutes with the action of $H$. Most of the times, we will consider decorations $\deco$ that are completely symmetric, i.e. such that all decorated edges $\decoedge_\ri{}$ are $\perm{}^\ri{}-$symmetric. In such cases the space of interactions fixed by the action of the whole group is again invariant under $\reno^\deco$ and we can focus on the action of the renormalization group on this smaller submanifold that is again going to be an embedded multiprojective space.\\[3pt]
We are now going to present, for each $\ri{}$, a decomposition of $\bw^\ri{}$ into subspaces that are invariant under $\perm{q}$; we will then select a \emph{fixed} vector in each of such subspaces and the set of such vectors will ultimately form a basis for the linear subspace of fixed vectors, that projected on $\bwp^\ri{}$ will give an embedded projective space. The same decomposition, applied to each factor of $\dynspace$, will give an embedded multiprojective space.  The same idea will then be used to find the appropriate multiprojective space in the case of $\perm{q-1}$, i.e. of an external magnetic field.\\[3pt]
We first need to classify basic invariant subspaces; in order to do so we need to define a variation of Young tableaux:
\begin{definition}
A \emph{Young diagram} represents a way to write a natural number $r$ as the sum of $k$ naturals $l_1\geq l_2\geq\cdots \geq l_k>0$. It is pictured as $r$ boxes arranged in $k$ rows as in the following example:
\[
\yng(4,2,1) \qquad 7=4+2+1.
\]
A (generalized) \emph{Young tableau} is a Young diagram in which we fill the boxes with numbers from 1 to $r$ according to the rule that numbers on the same row are increasing from left to right and numbers on the first column of rows of equal length are increasing from top to bottom, for example:
\[
\young(15,34,2)\ \mathrm{is\ OK,\ but}\ \young(34,15,2)\ \mathrm{is\ not.}
\] 
\end{definition}
This is \emph{not} the usual definition of Young tableaux involved in the classification of representation of the permutation group: in fact, for this purpose, each column would be ordered so as to be increasing from top to bottom as well. The definition we presented is, however, exactly what we need to classify basic invariant subspaces.\\
For each $\ri{}=(r,i)$, numbers from 1 to $r$ are associated to the corresponding spin of each $r-$tuple of vertices belonging to the edge set $E_\ri{}$; to each Young tableau with $r$ boxes and at most $q$ rows we associate the invariant subspace given by the following constraints: spins belonging to the same row have to be in the same state; spins belonging to different rows have to be in different states. In the case of completely symmetric decorations we can do the same with Young diagrams, as we can forget about the ordering of the spins.
For each invariant subspace there exists a one-dimensional space on which the permutations act trivially, that is the subspace generated by the sum of all base vectors; such vector will be denoted by $z$ with the corresponding Young tableau as a subscript; the direct sum of all such fixed spaces is obviously fixed by the permutation groups and it projects onto a projective space on $\bwp^\ri{}$.
\begin{example}
\label{ezexample}
Consider the case $\ri{}=(3,i),\ q=3$. The complex space of Boltzmann weights $\bw^\ri{}$ is a linear space of complex dimension 27 and it will have as a basis: 
\[
\begin{array}{ccccccccc}
e_{111}& e_{121} & e_{131} & e_{211}& e_{221} & e_{231} & e_{311}& e_{321} & e_{331}\\ 
e_{112}& e_{122} & e_{132} & e_{212}& e_{222} & e_{232} & e_{312}& e_{322} & e_{332}\\ 
e_{113}& e_{123} & e_{133} & e_{213}& e_{223} & e_{233} & e_{313}& e_{323} & e_{333}\\ 
\end{array}
\]
 All possible Young tableaux according to our definition, with the corresponding invariant subspaces are:
\begin{small}
\[
\begin{array}{rcl}
\young(123)&\rightarrow&\langle e_{111},e_{222},e_{333}\rangle\\[5pt]
\young(12,3)&\rightarrow& \langle e_{112},e_{113},e_{221},e_{223},e_{331},e_{332}\rangle\\[15pt]
\young(13,2)&\rightarrow& \langle e_{121},e_{131},e_{212},e_{232},e_{313},e_{323}\rangle\\[15pt]
\young(23,1)&\rightarrow& \langle e_{211},e_{311},e_{122},e_{322},e_{133},e_{233}\rangle\\[15pt]
\young(1,2,3)&\rightarrow& \langle e_{123},e_{132},e_{213},e_{231},e_{312},e_{321}\rangle
\end{array}
\]
\end{small}
where we denote by $\langle v_1,\cdots,v_k \rangle$ the $k-$dimensional complex vector subspace of $\cplx^{27}$ obtained by taking $\cplx-$linear combinations of the vectors $v_1,\cdots,v_k$.
The complex 1-dimensional fixed subspace associated to each tableau is generated by the sum of the corresponding base vectors.
\[
\begin{array}{rcl}
e_{\tiny\young(123)}&\defeq&e_{111}+e_{222}+e_{333}\\
e_{\tiny\young(12,3)}&\defeq&e_{112}+e_{113}+e_{221}+e_{223}+e_{331}+e_{332}\\
e_{\tiny\young(13,2)}&\defeq&e_{121}+e_{131}+e_{212}+e_{232}+e_{313}+e_{323}\\
e_{\tiny\young(23,1)}&\defeq&e_{211}+e_{311}+e_{122}+e_{322}+e_{133}+e_{233}\\
e_{\tiny\young(1,2,3)}&\defeq&e_{123}+e_{132}+e_{213}+e_{231}+e_{312}+e_{321}
\end{array}
\]
Passing to the quotient, this subspace of complex dimension 5 will therefore project down on $\bwp^\ri{}=\proj^{26}$ as an embedded $\proj^4$. \\
In the completely symmetric case, we can use Young diagrams instead of Young tableaux, obtaining a yet lower dimensional subspace, as the three subspaces corresponding to the Young diagram \begin{tiny}$\yng(2,1)$\end{tiny} are now part of the same subspace. Passing to the quotient we thus obtain an embedded $\proj^2$.
\end{example}
In the case of external magnetic field we will need to consider special Young diagrams and tableaux with a privileged row that do not mix under permutations with the others. This leads to even more complicated Young tableaux; in the following example we will consider completely symmetric decorations, so we can just use marked Young diagrams:
\begin{example}
Case $\ri{}=(2,i)$, $q=3$. We will consider state 1 as the special (magnetic) one. 
A natural basis for the complex space is:
\[
\begin{array}{ccc}
e_{11}& e_{12} & e_{13}\\
e_{21}& e_{22}& e_{23}\\
e_{31} & e_{32}& e_{33}\\ 
\end{array}
\]
 All possible marked Young diagrams, with the corresponding invariant subspaces are:
\Yboxdim{8.0pt}\Ylinethick{0.25pt}
\begin{small}
\[
\begin{array}{rcl}
\young(\gstar\gstar)&\rightarrow&\langle e_{11}\rangle\\[5pt]
\young(\gstar,\hfil)&\rightarrow& \langle e_{12},e_{13},e_{21},e_{31}\rangle\\[15pt]
\yng(2)&\rightarrow& \langle e_{22},e_{33}\rangle\\[15pt]
\yng(1,1)&\rightarrow& \langle e_{23},e_{32}\rangle\\[15pt]
\end{array}
\]
\end{small}
The projective space associated to this symmetry is therefore a $\proj^3\subset\bwp^\ri{}=\proj^8$.
\end{example}
\section{Physical variables}\label{physical}
\begin{figure}[!ht]
\centering{
\includegraphics[width=3cm]{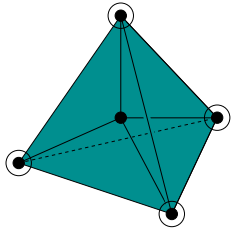}}
\caption{Decorated 4-edge analogous to the 3-edge in \fref{spiderweb}. A tetrahedron is split into four smaller tetrahedrons that join each of the four faces of the original one with its barycentre.}
\label{tistuta}
\end{figure}

\Yboxdim{3.0pt}\Ylinethick{0.25pt} 
In the previous section we presented the structure of the space $\dynspace$ on which the renormalization map acts; the space $\dynspace$ contains all multiple-spin interactions that can possibly be generated by the renormalization procedure and, as such, it is the natural space to consider for studying the dynamics of the renormalization map. However, from the physical point of view, we are usually interested in a restricted set of interactions, given, for instance, by pair-interaction between spins and coupling with an external magnetic field.\\ 
Following the reasoning in the previous section, we expect that this space, that we call \emph{physical space} and we denote by $\physspace$, can be given a natural structure of a product $\proj^1\times\proj^1$. In fact, defining a pair interaction amounts to assign a certain energy $\Jp$ to two neighbouring spins that are in the same state (parallel) and energy $\Jnp$ to the configuration for which they are in different states (antiparallel). The two values $\Jp$ and $\Jnp$ are affected by the arbitrary choice of zero of energy, thus, once more, we can define an equivalence relation on Boltzmann weights $(\zp,\znp)$. Equivalence classes are given in homogeneous coordinates by $[\zp:\znp]=[z:w]\in\proj^1$. The coupling with an external magnetic field can be treated in the same way: on a Potts model we choose a special state to be coupled to the field with energy $H_{\young(\star)}$ while all other states will have energy $H_{\yng(1)}$; these values are again affected by the choice of zero of energy so that we have another projective pair on Boltzmann weights, that we denote in the usual homogeneous coordinates by $[h_{\young(\star)}:h_{\yng(1)}]$.\\ 
For a given hierarchical lattice, one has to define how the physical space $\physspace$ is mapped into the dynamical space $\dynspace$. We will now present a canonical (and natural) way to embed the magnetic field variables in $\dynspace$. 
Let $\bGamma_\infty$ be a hierarchical lattice defined by iterating a decoration procedure $\deco$ on an initial hypergraph $\bGamma_0$. We introduce in $\dynspace^{\bGamma_\infty}$ an auxiliary space of 1-interactions $\proj^1$, given by the magnetic field variables  $[h_{\young(\star)}:h_{\yng(1)}]$; let $\tilde\dynspace^{\bGamma_\infty}=\dynspace^{\bGamma_\infty}\times\proj^1$. For each decorated edge $\decoedge$ of the decoration $\deco$, we attach to each \emph{core} vertex one 1-edge corresponding to the magnetic field variables; the auxiliary 1-edges will be decorated with the identity edge; let the resulting decoration be $\tilde\deco$. Finally, attach to each vertex of $\bGamma_0$ an 1-edge corresponding to the magnetic field variables; let the resulting hypergraph be $\tilde{\bGamma}_0$ and let $\tilde{\bGamma}_\infty$ be the hierarchical lattice generated by iteration of the decoration $\tilde\deco$ on $\tilde{\bGamma}_0$. It is easy to check that $\tilde{\bGamma}_\infty$ will have \emph{one} auxiliary edge attached to each vertex, therefore the magnetic field variables will induce a genuine coupling with an external magnetic field. It is important to note that since the auxiliary edges are not decorated, the external magnetic field variables will act as parameters of the renormalization map instead of being genuine dynamical variables. 
Recall that, in case of a magnetic field, one also has to take into account also a restricted symmetry of the states, as shown in the following example 
\begin{example}\label{campo}
Let us consider 2-interactions with a magnetic field. The dynamical space is $\proj^3\times\proj^1$; with homogeneous coordinates given by:
\[
\left[z_{\young(\star\star)}:z_{\young(\star,\hfil)}:z_{\yng(2)}:z_{\yng(1,1)}\right],\left[h_{\young(\star)}:h_{\yng(1)}\right]
\]
The natural embedding is:
\[
\left[z:w\right],\left[h_{\young(\star)}:h_{\yng(1)}\right]\mapsto\left[z_{\young(\star\star)}=z:z_{\young(\star,\hfil)}=w:z_{\yng(2)}=z:z_{\yng(1,1)}=w\right],\left[h_{\young(\star)}:h_{\yng(1)}\right]
\]
\end{example}
The situation for the pair-interaction variables is quite different, as we cannot define a canonical embedding of the pair-interaction variables as we did for magnetic field variables. In fact, the embedding depends on the particular hierarchical lattice we want to consider. In the following examples we present a number of cases.
\begin{example}
The easiest situation is given by a model on a completely symmetric 2-uniform hypergraph (i.e. a standard graph). In this case one maps directly the physical $\proj^1$ in the dynamical $\proj^1$ with the identity map:
\[
[z:w]\mapsto\left[z_{\yng(2)}=z:z_{\yng(1,1)}=w\right]. 
\]
\label{trivial} 
\end{example}
\begin{example}\label{tri}
Consider a model on a completely symmetric 3-uniform hypergraph without external magnetic field and $q\geq 3$. As stated in the previous section, example \ref{ezexample}, the dynamical space is a $\proj^2$. Suppose we want to put 2-spin interactions along each side of the triangle. This is a way to embed the projective pair $[z:w]$ in the dynamical space:
\[
[z:w]\mapsto\left[z_{\yng(3)}=z^3:z_{\yng(2,1)}=zw^2:z_{\yng(1,1,1)}=w^3\right].
\]
In fact, if all three spins are in the same state we have three parallel pairs i.e. $z_{\yng(3)}=z^3$; if two spins are in the same state and the third one is in a different state, then we have one parallel pair and a two antiparallel pairs i.e. $z_{\yng(2,1)}=zw^2$; finally if all three spins are in different states, then all pairs will be antiparallel i.e. $:z_{\yng(1,1,1)}=w^3$.\\
\end{example}
Notice that with the embedding defined in example \ref{tri}, each side will be counted as many times as the number of 3-edges that share that side. Sometimes this is undesirable, since such number can vary from side to side. In such cases one can add to the decorated edge some auxiliary 2-edges that will not be decorated (exactly as we did in the case of magnetic field variables) and that will be the edges carrying the physical pair-interaction. This formally adds to the dynamical space a new $\proj^1$ factor; again, since the auxiliary 2-edges are not decorated, interactions belonging to this $\proj^1$ will be considered as a parameter of the renormalization map. 
\begin{example}
Consider the decorated edge in \fref{tistuta}; at the $n$-th iteration each side of the original tetrahedron will be shared by $2^n$ 4-edges. If we want to avoid counting such multiplicities, we need to attach to the decoration four additional auxiliary 2-edges, namely the four sides that are \emph{inside} the tetrahedron. These 2-edges will not be decorated, but they will be the ones carrying the pair interaction of the physical space as in example \ref{trivial}; the dynamical variables associated to such edges will therefore act as parameters in the renormalization map. The dynamical space will be given by $\dynspace=\proj^4\times\proj^1$ and the embedding in this case is:
\[\fl
[z:w]\mapsto\left[z_{\yng(4)}=1:z_{\yng(3,1)}=1:z_{\yng(2,2)}=1:z_{\yng(2,1,1)}=1:z_{\yng(1,1,1,1)}=1\right],\left[z_{\yng(2)}=z:z_{\yng(1,1)}=w\right].
\]
\label{ingiuria}
\end{example}
Although the embedding of example \ref{ingiuria} is constant in the $\proj^4$ factor of $\dynspace$, the renormalization will create 4-edge interactions that will be carried by variables belonging to this factor.\\
As the examples suggest, the physical space $\physspace$ is mapped into the dynamical space (possibly after extending $\dynspace$ with new auxiliary interactions) as a submanifold; in general, this submanifold is not preserved by the dynamics of the renormalization map. This amounts to the well-known fact that the renormalization of pair interactions introduce, in general, new multiple-spin interactions. In any case, once we obtain all thermodynamical functions in the (possibly extended) dynamical space, it is easy to restrict to the physical space to obtain thermodynamical functions in \emph{relevant} coordinates. 
\section{The Green current and the set of zeros of the partition function}\label{main} 
As we showed, the generator of the renormalization map for hierarchical lattices can be represented by a rational map on a complex multiprojective space. We refer the interested reader to the appendices for a minimal technical introduction on the subject of iteration of such maps. The key result we are going to use is that a rational map comes quite naturally associated with a so-called \emph{Green current}, that can be thought as a differential form with distributional coefficients with support on the unstable set of the map. Such current is the limit under pull-back of the standard K\"ahler form if the map satisfies two properties called \emph{dominance} and  \emph{algebraic stability}.
Our goal is to show a connection between the Green current of the renormalization map and the non-analyticity locus of the free energy of the hierarchical lattice generated by the corresponding decoration. To prove such connection we use results that so far are only available for rational maps acting on projective spaces (not multi-projective spaces); for this reason, in what follows, we will consider only uniform hypergraphs and decorations, for which the renormalization map is acting on a projective space, although everything (but 
Theorem \ref{dinhsib}!) holds true in the more general setting.\\[3pt] 
Let us fix a $\ri{}-$uniform decorated edge $\decoedge$ and let $\deco$ be the decoration induced by $\decoedge$. The renormalization map is $\reno^\deco:\bwp^{\ri{}}\to \bwp^{\ri{}}$; let $d$ denote the algebraic degree of $\reno^\deco$, i.e. the degree of the polynomials we obtain lifting the map to $\bw^\ri{}$.\\ Fix now a $\ri{}-$uniform hypergraph ${\bGamma}_0$ and consider the zero set of the partition function $\fpart^{\deco^n{\bGamma}_0}$ of the $n$ times decorated hypergraph $\deco^n{\bGamma}_0$. By \eref{unicum}, this set is just the $n$-th preimage of the zero set of  $\fpart^{{\bGamma}_0}$ under the renormalization map. Such zero set is a codimension 1 algebraic variety that we will denote $LY_n$. If we consider the (normalized) current of integration $[LY_n]$ on the variety $LY_n$ we can express its relation to the current of integration $[LY_0]$ on the zeros $LY_0$ associated to ${\bGamma}_0$ in the following way:
\[
[LY_n]=\frac{1}{d^n}\left(\left(\reno^\deco\right)^n\right)^*[LY_0].
\]
Recall that the number of edges of the hypergraph $\deco^n{\bGamma}_0$ is $d^n$ times the number of edges of ${\bGamma}_0$; as $\fpart^{\deco^n{\bGamma}_0}=\fpart^{{\bGamma}_0}\circ\left(\reno^\deco\right)^n$, the free energy per edge of $\deco^n{\bGamma}_0$ is:
\[
\enel_{\deco^n{\bGamma}_0}=\frac{1}{\deg\fpart^{{\bGamma}_0}}\frac{1}{d^n}\log \left| \fpart^{{\bGamma}_0}\circ\left(\reno^\deco\right)^n \right|.
\]
The last formula shows that the free energy $\enel$ is just the pluripotential of the current supported on the zero locus of the polynomial $\fpart^{{\bGamma}_0}\circ{\reno^\deco}^n$. In the limit $n\rightarrow\infty$ the support of this current coincides with the Lee-Yang \cite{yl1,yl2} and Fisher zero locus of the model on the hierarchical lattice ${\bGamma}_\infty$. 
Results for this kind of limits have been found by Brolin \cite{br}, Lyubich \cite{ly} for \smash{$\proj^1$} in the 80s, by Favre-Jonnson \cite{fj} for holomorphic maps of \smash{$\proj^2$} in 2003. Very recently Dinh and Sibony proved the following
\begin{theorem}[Dinh-Sibony \cite{ds}]
Let $f\in\holo{d}(\proj^k)$ a holomorphic map of degree $d$ on the projective space of complex dimension $k$, $T$ its Green current. There exists a completely invariant proper analytic subset $E$ such that if $H$ is a hypersurface of degree $s$ in $\proj^k$ which does not contain any component of $E$, then
\[
\frac{1}{d^n}{f^n}^* [H] \to sT 
\] 
where $[H]$ is the current of integration on $H$. \label{dinhsib}
\end{theorem}
The maximal completely invariant proper subset $\mathscr{E}\supset E$ has been found (\cite{bcs}) to be a finite union of linear subspaces and bounds have been found for the maximal number of components of codimension 1 (\cite{fs}) that cannot be more than $k+1$ (sharp) and for codimension 2 (\cite{ac}) that is less than $4(k+1)^2$ (possibly not sharp).\\
We recall (see \ref{appA} for details) that, while rational maps on $\hat\cplx$ are automatically holomorphic, this is not true in general for rational maps in higher dimensional spaces; in fact, holomorphic maps are such that the so-called \emph{indeterminacy set} is empty. From the physical point of view, the indeterminacy set contains all Boltzmann weights that cannot be renormalized, i.e. such that applying the renormalization map to them gives all Boltzmann weights equal to 0. Although renormalization maps are not in general holomorphic, their restrictions on symmetrical interaction submanifold (see \sref{dynproj}) usually are. Moreover, the requirement of being holomorphic is a technical assumption that can possibly be removed using a more careful definition of the Green current.  
The connection is nevertheless interesting and it is worthwhile to try to understand how properties of the decoration are related to regularity properties of the corresponding renormalization map.
As summarized in the appendix, we need the map to enjoy two main properties in order for the Green current to be at least defined: \emph{dominance} and \emph{algebraic stability}. \\
The dominance property states that the Jacobian determinant of the map should not be identically zero. It is therefore very easy to check if a particular renormalization map enjoy this property; nevertheless it is interesting to point out that, in general, decoration that present some degeneracies will correspond to non-dominant maps. We now give two examples of such degenerate decorations:
\begin{example}
As a first example consider a decoration such that the renormalization map is \emph{invariant}  under permutations of $\perm{}^\ri{}$; a 2-decoration suffices to illustrate the fact:
\begin{center}
\begin{minipage}{5cm}
\begin{center}
\includegraphics[width=4cm]{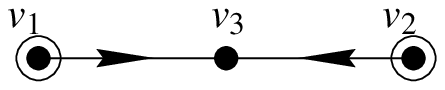}
\end{center}
\end{minipage}
\begin{minipage}{5cm}
\[
\fpart_{s_1s_2}=\fpart_{s_2s_1}.
\]
\end{minipage}
\end{center} 
 This implies that the image of the map is an algebraic subvariety, that in turn implies that the map is not dominant. This degeneracy is in some sense \emph{removable} as it can be ruled out by naturally restricting the map to the invariant variety which corresponds to $\perm{}^\ri{}-$invariant interactions.
\end{example}
\begin{example}
As a second example consider the following uniform decorated edge:
\begin{center}
\begin{minipage}{5cm}
\begin{center}
\includegraphics[width=2.5cm]{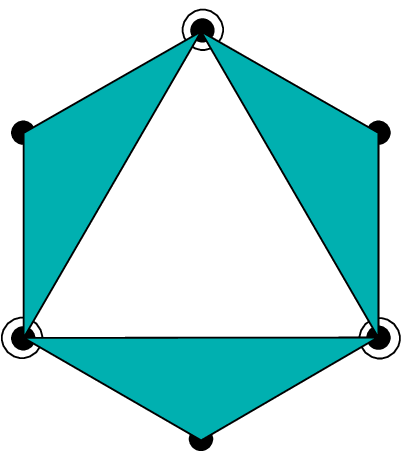}
\end{center}
\end{minipage}
\begin{minipage}{5cm}
\Yboxdim{3.0pt}\Ylinethick{0.25pt}
\[
\fpart_{\tiny\yng(3)}\cdot{\fpart_{\tiny\yng(1,1,1)}}^2 ={\fpart_{\tiny\yng(2,1)}}^3
\]
\end{minipage}
\end{center} 
In this case the 3-spin interactions can be expressed in terms of 2-spin interactions. Clearly the map will not be surjective on the space of 3-spin interactions as it will provide just interactions that can be described by $2-$edges, which in turn form a subvariety of codimension 1. 
In such cases one should again restrict to the appropriate space of interactions to obtain a dominant map.
\end{example}
The other regularity condition we have to check is \emph{algebraic stability}; this property is much harder to verify than the dominance condition. In fact, algebraic stability is related to the growth of the degrees of iterates of the renormalization map. It may happen that iterating the map we obtain factors that are common to all coordinates and which therefore have to be simplified; this operation lowers the degree of the map. In the maps studied so far, common factors do appear, but only in the definition of the map (i.e. the first iteration); we believe that once we simplify common factors which are possibly present at the first iteration, the renormalization map should be algebraically stable. Also, from a mathematical point of view, it would be quite important to prove algebraic stability for such maps, or at least to find conditions in terms of the decorations in order to ensure that this property holds. In fact, a characterization of algebraically stable maps is still lacking; for instance, it is not yet known how to build nontrivial maps that are a priori algebraically stable.       
\begin{example}
To give an example of the appearance of common factors we consider the model shown in figure \ref{cayley}. The model can be given by a non-uniform decoration; in this decoration we have two different kinds of one-dimensional edges (dotted and solid in the picture). The resulting graph is also known as the Cayley graph of the free group on 2 generators.\begin{figure}[h!]
\centering
\begin{minipage}{3cm}
\includegraphics[width=3cm]{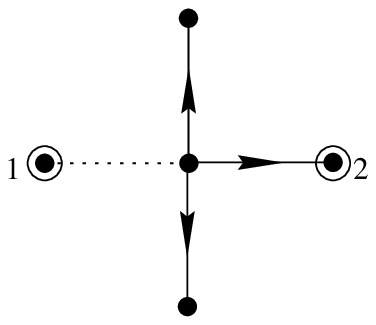}
\end{minipage}
\begin{minipage}{7cm}
\includegraphics[height=2.5cm]{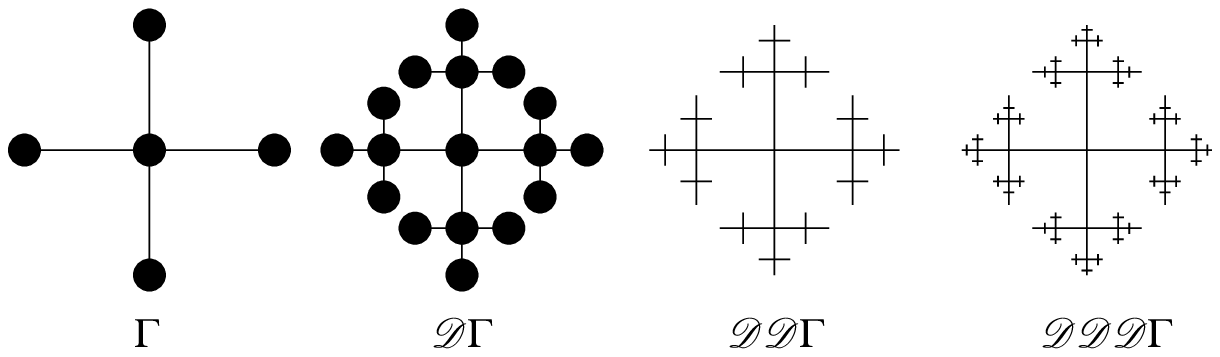}
\end{minipage}
\caption{Decoration generating the Cayley graph for the free group with two generators along with some iterations. This is not the simplest decoration that generates the Cayley graph, although this particular one have a renormalization map with common factors.}
\label{cayley} 
\end{figure}   
\end{example}
If we are in the case without an external magnetic field, the renormalization map associated to the decoration has common factors. Removing them corresponds to pruning all the branches of the tree and leaving a one-dimensional chain; this equivalence was observed long ago in \cite{egg}. This is the physical meaning to the idea of factoring out common factors in such a model although one probably cannot always give such a physical interpretation to the mathematical operation. 
\ack
The authors are indebted to C. Favre, S. Caracciolo and to anonymous referees for their most useful comments and suggestions.
\appendix
\section{Sum of decorated $\ri{}-$edges}
\label{sum}
In this appendix we define a natural sum operation on decorated edges. 
Let $\ri{}=(r,i)$ and let $\decoedge_1$ and $\decoedge_2$ be two decorated $\ri{}$-edges:
\begin{eqnarray*}
\decoedge_1&=&\{V=(v_1,\cdots,v_r)\sqcup V_0,E=E_\rib{1}\sqcup\cdots\sqcup E_\rib{p}\};\\\decoedge_2&=&\{W=(w_1,\cdots,w_r)\sqcup W_0,F=F_\ric{1}\sqcup\cdots\sqcup F_\ric{q}\};
\end{eqnarray*}
we define their sum $\decoedge_1 + \decoedge_2$ to be the decorated $\ri{}-$edge obtained by taking the disjoint union of the respective vertex and edge sets and then identifying surface vertices. The partition of the resulting edge set will be given by the union of the partitions of the summands: more formally let $\tilde V=(\tilde v_1,\cdots,\tilde v_r)\sqcup V_0\sqcup W_0.$ 
We define the collapsing map:
\[
\pi:V\sqcup W\to\tilde V,\]\[\pi(u)=
\cases{\tilde v_k&if $u=v_k$ or $u=w_k$ for some $k$\\u&otherwise i.e. $u\in V_0\sqcup W_0$\\},
\]
then the edge sets are given by:
\[
\tilde E_\rid{}\defeq \pi_* (E_\rid{}\sqcup F_\rid{})\quad \rid{}\in\bbeta\cup\bgamma,\ 
\]
and the sum decorated edge will be:
\[
\decoedge_1+\decoedge_2\defeq\{\tilde V,\tilde E\}.
\]
We can define the \emph{zero} decorated $\ri{}-$edge as the decorated $\ri{}$-edge with $r$ surface vertices, no core vertices and no edges; we consider the zero decorated edge to be uniform.
\[0_\ri{}\defeq\{V=(v_1,\cdots,v_r)\sqcup\emptyset,E=\emptyset\}.\]
Clearly the zero decorated edge is the null element of the sum operation.
It is straightforward to check that conditional partition function $\fpart^{0_\ri{}}_I$ of the zero decorated edge $0_\ri{}$ is constant. Also, it is easy to check that given $\decoedge_1$ and $\decoedge_2$ two uniform decorated $\ri{}-$edges we have that the renormalization map induced by the sum $\decoedge_1+\decoedge_2$ is given by the following expression 
\[
\reno^{\decoedge_1 + \decoedge_2}=\reno^{\decoedge_1}\cdot \reno^{\decoedge_2}
\]
where on the right hand side the product is defined coordinate-wise.
\section{Pluripotential theory}
\label{pluripotentialtheory}
In this appendix we give some basic notions about pluripotential theory which are useful in the study of the dynamics of the RG action. We refer the interested reader to the appropriate sections of \cite{sb, sib} for a more in-depth introduction.\\
Let $M$ be a smooth manifold and $\mathcal{D}(M)$ the vector space of smooth real-valued functions with compact support on $M$, endowed with the usual compact-open topology. The space of \emph{distributions}  $\mathcal{D}'(M)$ is the vector space of continuous linear functional on $\mathcal{D}(M)$ endowed with the usual weak topology.\\
Let $\Delta$ be the Laplace operator in $\cplx$ (as the 2-dimensional real Euclidean space); given a measure $\mu$ we define its \emph{potential} as the distributional solution of the equation $\Delta P_\mu=\mu$.
Functions that are local potentials of a positive measure $\mu$ are called \emph{subharmonic} and are characterized as follows:  
\begin{appdefinition}
Let $\Omega$ be an open domain of $\cplx$. An upper semi-continuous function $u:\Omega\to[-\infty,+\infty[$ is \emph{subharmonic} if it is not identically equal to $-\infty$ and it enjoys the subaverage property i.e. for all $z_0\in\Omega$, for all $r\in\reals^+$ such that the closed disk of center  $z_0$ and radius $r$ is contained in $\Omega$, we have 
\[ u(z_0)\leq\frac{1}{2\pi}\int_0^{2\pi}u\left(z_0+re^{i\theta}\right)\deh\theta  
\]
\end{appdefinition}
For example if $f$ is an holomorphic function then $u=\log |f|$ is subharmonic and $\Delta u$ is supported on the zeroes of $f$.\\[2pt]
In the multidimensional setting we will need to use \emph{currents} and \emph{plurisubharmonic functions} rather than distributions and subharmonic functions. We will now introduce the appropriate definitions.\\
Let $\mathcal{D}^p$ be the vector space of smooth differential $p$-forms with compact support endowed with the compact-open topology. A current $S$ of dimension $p$ is a continuous linear functional on $\mathcal{D}^p$; the space of $p$-currents will be denoted as ${\mathcal{D}^p}'$ and will be given the weak topology.
For example, since one can associate the Dirac delta to a point, one can associate a $p$-current to any $p$-dimensional submanifold $N$ of $M$ by integrating $p$-forms over $N$. 
Operations on forms as exterior product with other forms and the exterior differential operator can act by duality on the space of currents as well:\[
\langle S\wedge\omega,\phi\rangle\defeq \langle S,\omega\wedge\phi\rangle\qquad\langle\deh S,\phi\rangle\defeq (-1)^{p+1}\langle S,\deh \phi\rangle
\]
As a dual object to forms, a current $S$ can naturally be pushed forward by a map $f$, provided that the restriction of $f$ to the support of $S$ is proper (i.e. the preimage of compact sets is compact). Moreover, if $f$ is a proper submersion one can define a push-forward operation for forms and therefore one can define a pull-back for currents. 
If the manifold has a complex structure we should distinguish between the holomorphic and antiholomorphic part of a form. A complex differential form of bidegree $(p,q)$ can be written as:
\[
\mathcal{D}^{p,q}\ni\phi=\sum_{|I|=p |J|=q} \phi_{IJ} \deh z_I\wedge \deh \bar{z}_J
\]
A $(p,p)$-form is said to be \emph{positive} if for all complex submanifold $Y$ of dimension $p$, its restriction to $Y$ is a nonnegative volume form; $(p,q)-$currents are defined by duality and a $(p,p)-$current is said to be positive if it evaluates as a positive number on any positive $(p,p)-$form.\\
Along with the exterior holomorphic $\partial$ and antiholomorphic $\bar{\partial}$ differentiation we can define two real operators $\deh=\partial+\bar\partial$ and $\deh^c=\frac{i}{2\pi}\left(\bar\partial-\partial\right)$. The second order operator $\deh\deh^c$ is going to replace the Laplacian operator in the multidimensional setting. We are now left to introduce the analogous of subharmonic functions. 
\begin{appdefinition}
Let $\Omega$ be an open subset of $\cplx^n$. An upper semi-continuous function $u:\Omega\to[-\infty,\infty[$ is \emph{plurisubharmonic} (in short \emph{psh}) in $\Omega$ if it is not identically equal to $-\infty$ and it enjoys the subaverage property when restricted to any 1-dimensional disk i.e. for all $z_0\in\Omega$ and for all $w\in\cplx^n$ such that the one-dimensional complex disk $z_0+w\bar\mathbb{D}$ (where $\bar\mathbb{D}$ is the closed unit disk in $\cplx$) is contained in $\Omega$ one has
\[ u(z_0)\leq\frac{1}{2\pi}\int_{0}^{2\pi}u\left(z_0+we^{i\theta}\right)\deh \theta
\]
\end{appdefinition}
The space of psh functions enjoys an important compactness property:
\begin{apptheorem}
Let $u_j$ be a sequence of plurisubharmonic functions on a domain $\Omega\subset\cplx^n$. Assume that for all compacts $K\subset\Omega$ the sequence is dominated by a psh function. Then either $u_j\to-\infty$ on all compact subsets of $\Omega$ or there exists a subsequence $u_{j_k}$ which converges in $L^1_{\mathrm{loc}}(\Omega)$ to a psh function.
\end{apptheorem}
A function $u\in L_\mathrm{loc}^1(\Omega)$ is a.e. equal to a psh function  the $(1,1)-$current $\deh \deh^c u$ is positive; conversely if $S$ is a positive closed $(1,1)-$current, there exists a psh function $u$ such that $u$ is a local potential of $S$.
\section{Projective spaces and rational dynamics}
\label{appA}
Consider the complex vector space $\cplx^{n+1}\setminus\{0\}$ modulo the action of the multiplicative group $\cplx^*$ by scalar multiplications. The resulting space is a complex manifold of dimension $n$ called projective space $\proj^n$. The natural coordinates on the projective space are the so called homogeneous coordinates:
\[
\proj^n\ni[z_0:z_1:\cdots:z_n]\defeq \pi(z_0,z_1,\cdots,z_n)
\]
where $\pi$ is the projection map that defines the quotient. $\proj^n$ comes naturally endowed with a standard K\"ahler form $\omega$ given by the relation $\pi^*\omega=\ma \log |z|$. 
\\
A rational map of degree $d$ over $\proj^n$ is a map of the form:
\[
f:[z_0:z_1:\cdots:z_n]\mapsto [P_0:P_1:\cdots:P_n]
\]
where $P_j$s are homogeneous polynomials of degree $d$ with no nonzero common factors. The map $f$ can be lifted to a polynomial map $F$ on the complex space up to nonzero multiplicative factors. A rational map on $\proj^n$ is said \emph{dominant} if given any lift $F$, its Jacobian determinant does not vanish identically. The set of dominant maps of degree $d$ will be denoted by $\dominanti{d}$. One then defines the \emph{indeterminacy set} $I\defeq\pi F^{-1}\left(\{0\}\right)$.\\
Roughly speaking $I$ is a \emph{bad} set for the dynamics and \emph{good} maps are such that $I$ is small. The space $\holo{d}\subset\dominanti{d}$ of maps such that $I=\emptyset$ is defined as the space of holomorphic maps. In most applications a weaker condition on $f \in \dominanti{d}$ suffices: suppose there is no integer $n$ and no codimension 1 hypersurface $V$ such that $f^n(V)\subset I$; then $f$ is said to be \emph{algebraically stable} as the latter condition is equivalent to require that the degree of $f^n$ is $d^n$.\\
A rational map $f$ acts on the space of positive closed $(1,1)-$currents by pull-back i.e. given a potential $u$ of a current $S$ (i.e. $\ma u=\pi^*S$), $f^*S$ is defined by the relation $\pi^* f^* S=\ma (u\circ F)$. Such action is continuous provided that $f$ is dominant. An important result is the following 
\begin{apptheorem}[see \cite{sib}]
Let $d\geq 2$ $f\in\dominanti{d}(\proj^N)$ algebraically stable. Then the sequence 
\[
T_n\defeq \frac{1}{d^n} \left(f^n\right)^*\omega
\]
converges to a closed positive $(1,1)-$current $T$ such that $f^* T=d\cdot T$. $T$ is called the \emph{Green current} of $f$. A potential of $T$ is called \emph{Green function}.
\end{apptheorem}
The support of the Green current can be partially understood in a purely topological setting; in fact, let us define the \emph{stable} (or Fatou) set of the map as follows:
\[ 
\fl
\fatou=\{p\in\proj^n \st \ex U\ni p \mathrm{\ open\ nbhd\ on\ which\ the\ family\ } f^k|_U \mathrm{\ is\ equicontinuous} \}\]
$\julia\defeq\proj^n\setminus\fatou$ is called \emph{Julia set} of $f$ and is the \emph{unstable} set for the dynamics; this set always contains the support of the Green current (see \cite{sib}), that therefore assume a definite topological meaning.\\[3pt]
A multiprojective space is just a product of $p$ projective spaces; rational maps on such spaces are those that are lifted to separately homogeneous polynomials. The notion of degree becomes that of multi-degree, that is a square integer matrix of dimension $p$. Studying the dynamics of rational maps on such spaces is more complicated and very few results have been proved so far (\cite{fg}), but among these there is the existence of the Green current for algebraically stable dominant maps. 

\end{document}